\documentclass[conference]{IEEEtran}
\IEEEoverridecommandlockouts
\usepackage{cite}
\usepackage{amsmath,amssymb,amsfonts}
\usepackage{algorithmic,algorithm}
\usepackage{graphicx}
\usepackage{textcomp}
\usepackage{xcolor}
\usepackage{ulem}
\def\BibTeX{{\rm B\kern-.05em{\sc i\kern-.025em b}\kern-.08em
    T\kern-.1667em\lower.7ex\hbox{E}\kern-.125emX}}
\usepackage{booktabs}

\definecolor{pli-color}{HTML}{002FA7}

\makeatletter
\newcommand{\linebreakand}{%
  \end{@IEEEauthorhalign}
  \hfill\mbox{}\par
  \mbox{}\hfill\begin{@IEEEauthorhalign}
}

\begin{document}

\title{Enhancing Quantum Security over Federated Learning via Post-Quantum Cryptography}

\author{\IEEEauthorblockN{Pingzhi Li}
\IEEEauthorblockA{\textit{Department of Computer Science} \\
\textit{The University of North Carolina at Chapel Hill}\\
Chapel Hill, United States \\
\texttt{pingzhi@cs.unc.edu}} 
\linebreakand
\IEEEauthorblockN{Tianlong Chen}
\IEEEauthorblockA{\textit{Department of Computer Science} \\
\textit{The University of North Carolina at Chapel Hill}\\
Chapel Hill, United States \\
\texttt{tianlong@cs.unc.edu}} \hspace{200pt}
\and
\IEEEauthorblockN{Junyu Liu}
\IEEEauthorblockA{\textit{Department of Computer Science} \\
\textit{The University of Pittsburgh} \\
Pittsburgh, United States\\
\texttt{junyuliu@pitt.edu}}
}

\maketitle

\begin{abstract}
% This document is a model and instructions for \LaTeX.
% This and the IEEEtran.cls file define the components of your paper [title, text, heads, etc.]. *CRITICAL: Do Not Use Symbols, Special Characters, Footnotes, 
% or Math in Paper Title or Abstract.

Federated learning~(FL) has become one of the standard approaches for deploying machine learning models on edge devices, where private training data are distributed across clients, and a shared model is learned by aggregating locally computed updates from each client. While this paradigm enhances communication efficiency by only requiring updates at the end of each training epoch, the transmitted model updates remain vulnerable to malicious tampering, posing risks to the integrity of the global model. Although current digital signature algorithms can protect these communicated model updates, they fail to ensure quantum security in the era of large-scale quantum computing. Fortunately, various post-quantum cryptography algorithms have been developed to address this vulnerability, especially the three NIST-standardized algorithms - Dilithium, FALCON, and SPHINCS+. 

In this work, we empirically investigate the impact of these three NIST-standardized PQC algorithms for digital signatures within the FL procedure, covering a wide range of models, tasks, and FL settings. Our results indicate that Dilithium stands out as the most efficient PQC algorithm for digital signature in federated learning. Additionally, we offer an in-depth discussion of the implications of our findings and potential directions for future research.

\end{abstract}


\section{Introduction}

Modern mobile devices have access to vast amounts of data that can be leveraged to train machine learning models, leading to significant improvements in the user experience~\cite{8994206, ZHANG2021106775}. For instance, language models can enhance speech recognition~\cite{7472820} and text entry~\cite{James2001TextIF} and function as intelligent assistants~\cite{8301638}. However, the data-driven nature of these models often requires continuous user data collection to refine and update the machine-learning models~\cite{9708956}. This challenge becomes even more complex in large-scale distributed systems, such as smartphone networks, which involve millions of users. Federated learning addresses this issue, emerging as a promising machine learning paradigm, where many devices (\textit{i.e.}, clients) collaborate to train a machine learning model while keeping the data on the devices themselves~\cite{mcmahan2023communicationefficientlearningdeepnetworks,ZHANG2021106775}. Various approaches have been developed to safeguard user privacy and ensure data security in the federated learning context.

\begin{figure}[tb]
    \centering
    \includegraphics[width=1.0\linewidth]{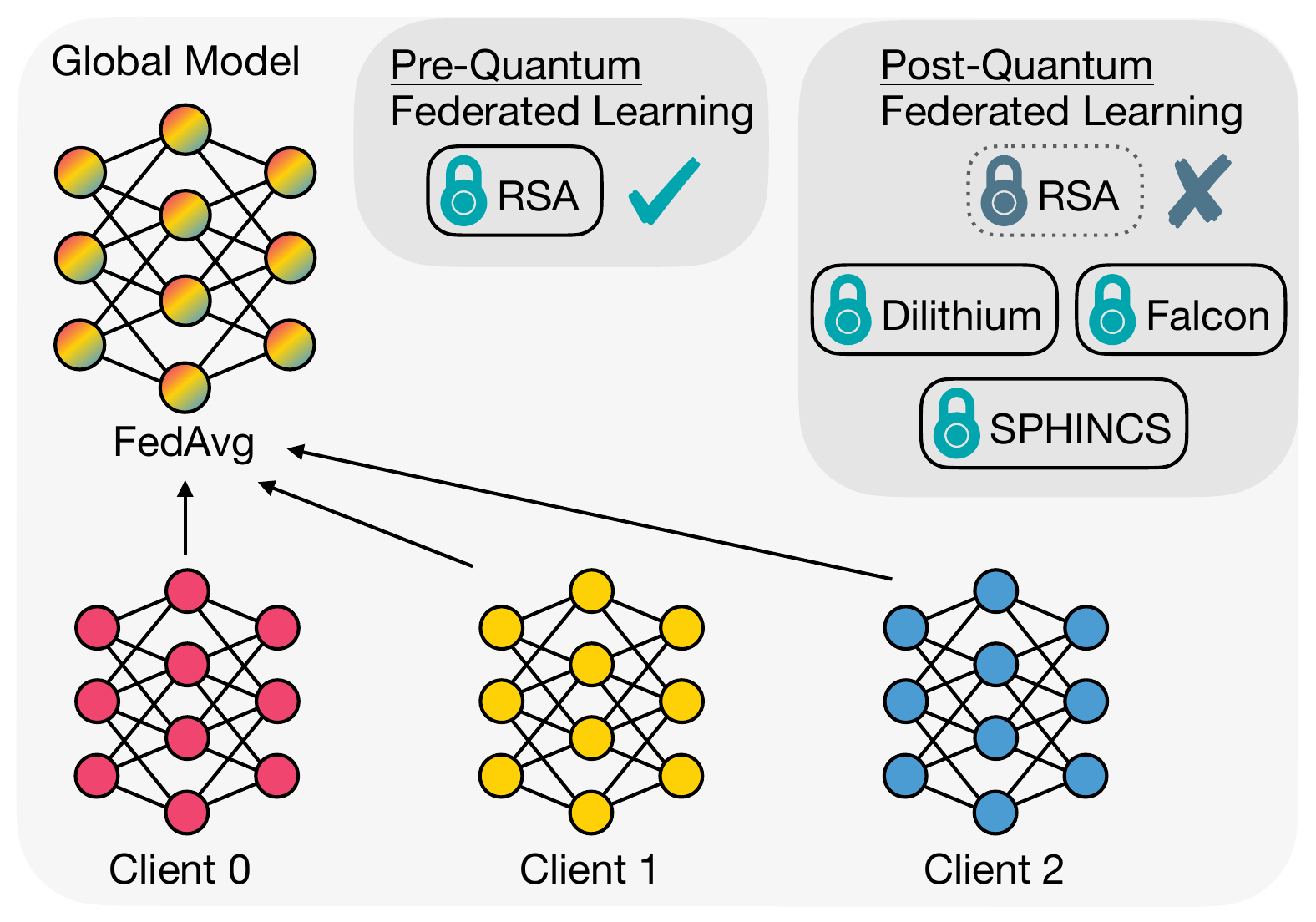}
    \caption{\textit{(Left)}~In the current pre-quantum federated learning landscape, the training process can be safeguarded against poisoning attacks using public-key cryptography algorithms like RSA. \textit{(Right)}~However, in the post-quantum era, classical cryptographic algorithms(\textit{e.g.}, RSA) are vulnerable to being broken by large-scale quantum computers and can no longer provide adequate security. Instead, post-quantum cryptography algorithms, such as Dilithium, Falcon, and SPHINCS+, should be employed for digital signature authentication.}
    \label{fig:intro-teaser}
\end{figure}

Among the various federated learning variants, one of the most widely used and foundational ones is \texttt{FedAvg}~\cite{mcmahan2023communicationefficientlearningdeepnetworks}. This approach distributes the training data across the clients and learns a shared model by aggregating locally computed updates for each client. During each epoch, every client computes an update to the current global model, which is maintained by a central server, and only this update is communicated at the end of the epoch. Although \texttt{FedAvg} is communication-efficient, the model updates transmitted from clients to the server can be vulnerable to malicious tampering, posing security threats to the global model. For example, attackers could train a separate language model on toxic content and substitute the federated learning language model update during a client device’s communication with this malicious update, thereby injecting toxicity into the global language model.

While current federated learning applications are safeguarded from communication-based threats through the use of digital signature algorithms that rely on public-key cryptography, they are not immune to quantum threats. Quantum computers, which have seen significant progress in recent years, utilize quantum mechanical principles to tackle mathematical problems beyond conventional computers' capabilities. Most widely used public-key algorithms rely on the difficulty of one of three mathematical problems: the integer factorization problem, the discrete logarithm problem, and the elliptic-curve discrete logarithm problem. For example, in the widely-used RSA~\cite{10.1145/359340.359342} public-key cryptosystem, the public key is represented as a product $N = pq$ of two secret prime numbers, $p$ and $q$. The security of RSA fundamentally depends on the computational difficulty of factorizing $N$ into its prime components, $p$ and $q$. However, in $1994$, Shor~\cite{Shor_1997} proposed a quantum algorithm capable of quickly determining the prime factorization of any positive integer $N$. If large-scale quantum computers are successfully realized, they can compromise many of today's public-key cryptosystems. This would pose a serious risk to the security and privacy of federated learning systems, compromising both their confidentiality and integrity.

Fortunately, significant progress has been made in Post-Quantum Cryptography~(PQC), which aims to develop cryptographic algorithms, such as public-key algorithms, that are believed to be resistant to cryptanalytic attacks by quantum computers. The NIST standardization process for PQC began in $2017$ with $69$ candidate algorithms for key establishment mechanisms~(KEM) and digital signature algorithms~(DSA). Subsequently, in $2019$, $2020$, $2021$, and $2022$, NIST conducted four additional rounds of evaluation, ultimately selecting $1$ PQC algorithm for KEM and $3$ PQC algorithms for DSA. Specifically, the $3$ DSA algorithms we plan to investigate in the context of federated learning are Dilithium, FALCON, and SPHINCS+.

In this paper, we empirically study the impact of $3$ PQC NIST standardized algorithms for DSA on federated learning, covering a range of model scales. Furthermore, we extend our study to quantum federated learning, where a quantum neural network is trained in the federated learning setting equipped with PQC DSA algorithms. Our study reveals the critical efficiency characteristics of those standard PQC DSA algorithms on federated learning, providing key insights for their application in the future. Our key contributions are summarized below:

\begin{itemize}
    \item We empirically study the impact of DSA algorithms in federated learning in the near future of the PQC era.
    \item Our comprehensive evaluation results show that Dilithium stands out as the most efficient PQC DSA algorithm in the context of federated learning poisoning attack defense. 
\end{itemize}

% todo: add a discussion about existing classical & quantum attack methods; add more discussion & figures.
% todo: potential limitation in discussion, what are the potential missing components.

\section{Threats to Federated Learning Security}

\subsection{Communication in Federated Learning}\label{sec:fl-communication}

\paragraph{Federated Learning Problem Definition} Consider a federated learning scenario involving a central server $\mathtt{G}$ and $M$ clients ${\mathtt{C}_{1\dots M}}$. Each client $\mathtt{C}_{i}$ possesses a private training dataset $\mathcal{D}_i=\{\mathtt{X}_i^j, \mathtt{Y}_i^j\}_{j}$, which remains accessible only to $\mathtt{C}_{i}$. The goal of the server $\mathtt{G}$ is to utilize the data distributed across all clients to learn a global model that minimizes the average loss across the entire dataset $[\mathcal{D}_1, \dots, \mathcal{D}_M]$, \textit{i.e.}, $\mathcal{L}=\frac{1}{M}\sum_{j=1}^M\texttt{CrossEntropy}(\mathtt{C}_j(\mathtt{X}_j), \mathtt{Y}_j)$. The most common method to address this problem is \texttt{FedAvg}.

\paragraph{\texttt{FedAvg} Communication} In each round $t$ of \texttt{FedAvg}, each client $\mathtt{C}_i$ receives the current global parameters $\theta_{\mathtt{G}}^t$ and performs multiple training steps on its local dataset $\mathcal{D}_i$, starting from $\theta_{\mathtt{G}}^t$. After completing these steps, the client sends the resulting model updates $\Delta\theta_{i}^t$ back to the server. The server then aggregates the updates from all clients into an average $\Delta\theta^t = \frac{1}{M}\sum_{i=1}^{M}\Delta\theta_i^t$ and uses this to update the global model: $\theta_{\mathtt{G}}^{t+1} = \theta_{\mathtt{G}}^{t} + \Delta\theta^t$. Thus, the only communication content in each round consists of the global model parameters $\theta_{\mathtt{G}}^t$, and each client's model updates $\Delta\theta_{i}^t$.

\begin{figure}[htb]
    \centering
    \includegraphics[width=1.0\linewidth]{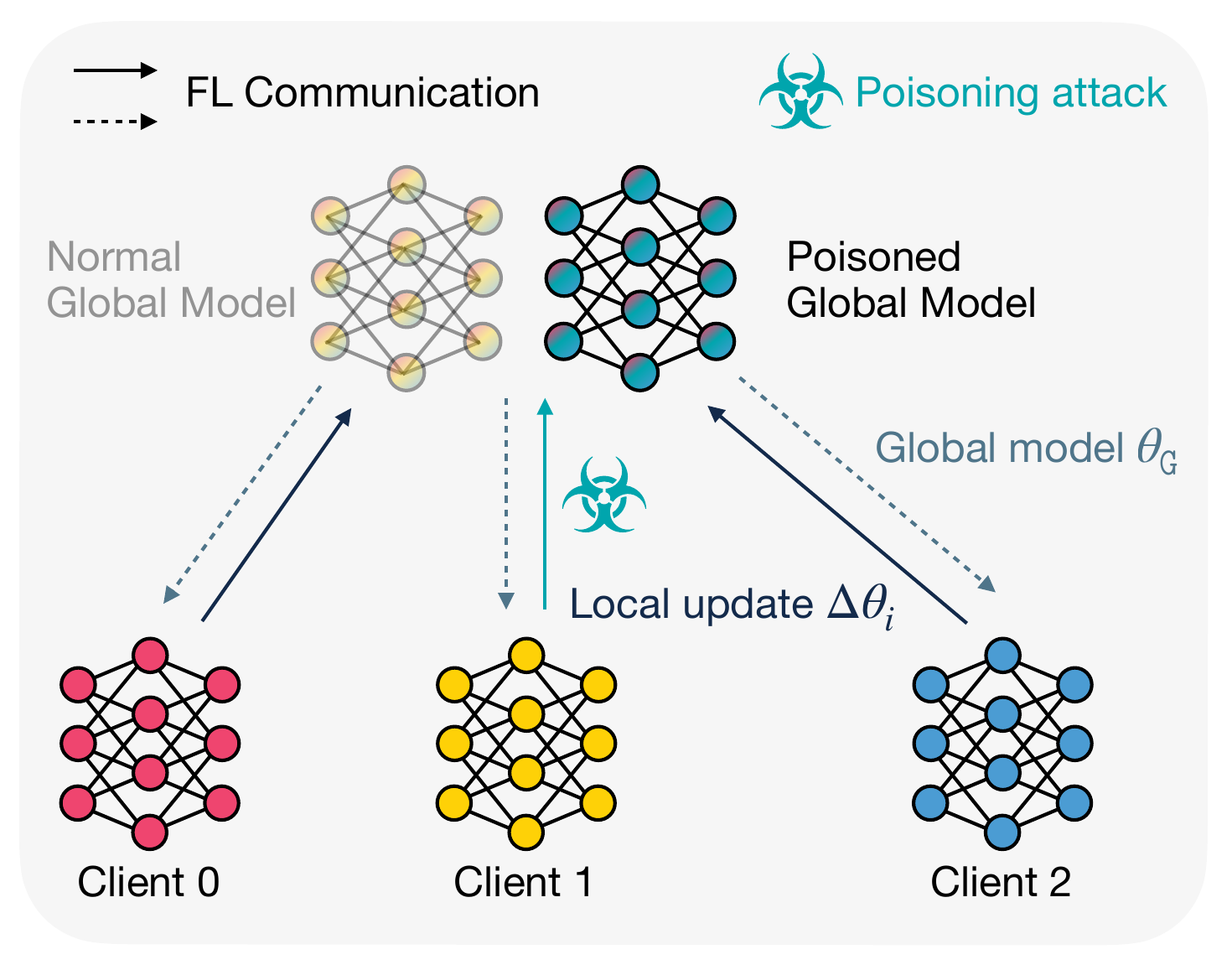}
    \caption{Poisoning attack from outsiders on the communicated model parameters in federated learning.}
    \label{fig:poisoning-attack}
\end{figure}

\subsection{Poisoning Attacks and Federated Learning}

Poisoning attacks in machine learning refer to the deliberate alteration of training data with the intent of causing the learned model to fail in detecting subsequent attacks~\cite{10.1145/1644893.1644895}. For instance, if the training dataset of a traffic sign recognition system is intentionally corrupted, it could lead autonomous vehicles that rely on the system to identify traffic signs~\cite{9019666} incorrectly. Generally, poisoning attacks can be classified into two categories: \uline{model failure} poisoning attacks, which aim to render the model unusable, and \uline{targeted error} poisoning attacks, which seek to induce the model to misclassify a specific label as another target label. 

In the context of federated learning, poisoning attackers can be categorized as insiders or outsiders. Insider attackers are local clients that maliciously contribute poisoned model updates, while outsiders compromise the communication between the central server and local clients, injecting poisoned parameters into the model updates. Existing studies primarily address insider attackers, given that outsider attackers can typically be mitigated using standard digital signature algorithms employed in modern networks. Our work extends this focus by addressing post-quantum security, where the classical digital signature algorithms are no longer effective, specifically aiming to defend against \uline{outsider poisoning attacks}, assuming that both the central server and local clients operate honestly. As shown in Figure~\ref{fig:poisoning-attack}, an outlier attacker poisons on the local update \textit{i.e.}~$\Delta\theta_{1}$ uploaded from client $1$, which will ultimately also poisons on the global model as it is updated based on every local model updates.

\section{Methodology}

In this section, we introduce the $3$ NIST-standardized PQC algorithms for digital signature and how we employ them in federated learning to enhance quantum security.

\subsection{NIST-Standardized Post-Quantum Cryptography}

In this part, we briefly introduce the $3$ NIST-standardized PQC algorithms, \textit{i.e.} Dilithium, Falcon, and SPHINCS+. We summarize the key characteristics of them as listed in Table~\ref{tab:algo-summary}.

\begin{table}[htbp]
  \centering
  \caption{Summary of the $3$ NIST-standardized PQC methods for digital signature. We marked the best ones for the key size, signature size, speed, and memory characteristics with \textbf{bold}.}
   \resizebox{\linewidth}{!}{
    \begin{tabular}{l|r|cccc}
    \toprule
    \midrule
    Method & Basis & Key Size & Signature Size & Speed & Memory \\
    \midrule
    Dilithium~\cite{CRYSTALS-Dilithium} & Lattice-based & Medium & Medium & \textbf{High} & Medium \\
    Falcon\footnote{https://falcon-sign.info/} & Lattice-based & Large & \textbf{Small} & Medium & Medium \\
    SPHINCS+~\cite{10.1007/978-3-662-46800-5_15} & Hash-based & \textbf{Small} & Large & Slow & \textbf{Low} \\
    \midrule
    \bottomrule
    \end{tabular}}
  \label{tab:algo-summary}
\end{table}

\paragraph{Dilithium} CRYSTALS-Dilithium~\cite{CRYSTALS-Dilithium} is a lattice-based digital signature scheme built using the Fiat-Shamir heuristic. Its security is based on the hardness of solving the Module Learning With Errors~(MLWE) and Module Short Integer Solution~(MSIS) problems. Dilithium offers a balanced performance in terms of key and signature size and efficiency in key generation, signing, and verification. It uses the ring $R_q := Z_q[X]/(X^256 + 1)$, where q is the prime number $2^23 - 2^13 + 1$. Dilithium has been implemented on various platforms, including ASIC, FPGA, and RISC-V, showing significant performance improvements over traditional implementations.

\paragraph{Falcon} Falcon\footnote{https://falcon-sign.info/} is another lattice-based digital signature scheme that follows a "hash and sign" paradigm. Its security is based on the difficulty of solving the Short Integer Solution~(SIS) problem on NTRU lattices. Falcon stands out for its compact signatures, smaller than Dilithium at comparable security levels, though its public key size is moderately larger. This characteristic makes Falcon particularly attractive for applications where signature bandwidth is a primary concern. While Falcon has more complex hardware implementation requirements than Dilithium due to its data tree structure and advanced mathematical operations, it demonstrates superior efficiency in signing and verifying operations. The scheme integrates well with existing protocols and has shown promising results in various implementations, including high-performance cryptographic processors.

\paragraph{SPHINCS+} SPHINCS+~\cite{10.1007/978-3-662-46800-5_15} is a stateless hash-based digital signature scheme, distinguishing it from the lattice-based approaches of Dilithium and Falcon. Its security is fundamentally based on the security of the underlying hash function, making it a conservative choice in the post-quantum landscape. SPHINCS+ has the largest signature size among the NIST candidates but compensates with the smallest public key size. It is relatively slow in execution compared to its lattice-based counterparts, particularly for signing operations. However, its stateless nature provides important advantages in certain applications. Various implementations of SPHINCS+ have been explored, including parallelized designs and FPGA implementations, showing potential for performance improvements despite its inherent computational intensity.

\subsection{FL Quantum Security via PQC}

To enhance the quantum security of federated learning (FL), we integrate PQC digital signature algorithms into the FL communication protocol. This integration aims to protect against potential quantum-enabled attacks on the model updates transmitted between clients and the central server. We implement this security enhancement using the three NIST-standardized PQC algorithms for digital signatures: Dilithium, Falcon, and SPHINCS+.
As defined in Section~\ref{sec:fl-communication}, let $\theta^t_\mathtt{G}$ represent the global model parameters at round $t$, and $\Delta\theta^t_i$ denote the model updates from client $i$. We modify the standard FL protocol as follows:

\begin{itemize}
    \item \textbf{Key Generation}: At the beginning of the FL process, each client $\mathtt{C}_i$ and the central server G generate their respective public-private key pairs using one of the PQC algorithms: $(\mathtt{pk}_i, \mathtt{sk}_i)$ for client $\mathtt{C}_i$, $(\mathtt{pk}_\mathtt{G}, \mathtt{sk}_\mathtt{G})$ for the central server $\mathtt{G}$.
    \item \textbf{Model Distribution}: When the server sends the global model $\theta^t_\mathtt{G}$ to client $\mathtt{C}_i$, it signs the message using its private key: $\sigma_\mathtt{G} = \texttt{sign}(\mathtt{sk}_\mathtt{G}, \theta^t_\mathtt{G})$. The client verifies the signature using the server's public key before accepting the model: $\texttt{verify}(\mathtt{pk}_\mathtt{G}, \theta^t_\mathtt{G}, \sigma_\mathtt{G})$.
    \item \textbf{Update Submission}: After computing local updates $\Delta\theta^t_i$, client $\mathtt{C}_i$ signs the update before sending it to the server: $\sigma_i = \texttt{sign}(\mathtt{sk}_i, \Delta\theta^t_i)$ The signed update $(\Delta\theta^t_i, \sigma_i)$ is then sent to the server.
    \item \textbf{Update Verification}: Upon receiving the signed update, the server verifies the signature using the client's public key: $\texttt{verify}(\mathtt{pk}_i, \Delta\theta^t_i, \sigma_i)$. Only verified updates are included in the aggregation process.
    \item \textbf{Model Aggregation}: The server aggregates the verified updates to compute the new global model: $\theta^{t+1}_\mathtt{G} = \theta^{t}_\mathtt{G} + (1/M) \Sigma_{i=1}^M \Delta\theta^t_i$.
\end{itemize}

We provide the formal algorithm description in Algorithm~\ref{alg:pqc-fl}. This protocol ensures that all communicated model parameters and updates are authenticated using quantum-resistant signatures, mitigating the risk of tampering by quantum-enabled adversaries. The choice of the PQC algorithm (Dilithium, Falcon, or SPHINCS+) affects the performance characteristics of this secure FL system, which we evaluate in our experiments.
By implementing this PQC-enhanced protocol, we provide a robust defense against potential quantum attacks on the FL communication process, ensuring the integrity and authenticity of model updates in a post-quantum environment.

\begin{algorithm}
\caption{PQC-Enhanced Federated Learning}
\label{alg:pqc-fl}
\begin{algorithmic}[1]
\STATE \textbf{Input:} Number of clients $M$, number of rounds $T$
\STATE \textbf{Output:} Secure global model $\theta^T_\mathtt{G}$

\STATE \textbf{Key Generation:}
\FOR{each client $\mathtt{C}_i$ and server $\mathtt{G}$}
    \STATE $(\mathtt{pk}_i, \mathtt{sk}_i) \leftarrow \texttt{keyGen()}$
    \STATE $(\mathtt{pk}_\mathtt{G}, \mathtt{sk}_\mathtt{G}) \leftarrow \texttt{keyGen()}$
\ENDFOR

\FOR{each round $t = 1$ to $T$}
    \STATE \textbf{Model Distribution:}
    \FOR{each client $\mathtt{C}_i$}
        \STATE $\sigma_\mathtt{G} \leftarrow \texttt{sign}(\mathtt{sk}_\mathtt{G}, \theta^t_\mathtt{G})$
        \STATE Send $(\theta^t_\mathtt{G}, \sigma_\mathtt{G})$ to client $\mathtt{C}_i$
        \STATE Client $\mathtt{C}_i$: $\texttt{verify}(\mathtt{pk}_\mathtt{G}, \theta^t_\mathtt{G}, \sigma_\mathtt{G})$
    \ENDFOR
    
    \STATE \textbf{Local Update:}
    \FOR{each client $\mathtt{C}_i$ in parallel}
        \STATE Compute local update $\Delta\theta^t_i$
        \STATE $\sigma_i \leftarrow \texttt{sign}(\mathtt{sk}_i, \Delta\theta^t_i)$
        \STATE Send $(\Delta\theta^t_i, \sigma_i)$ to server $\mathtt{G}$
    \ENDFOR
    
    \STATE \textbf{Update Verification and Aggregation:}
    \STATE $S \leftarrow \emptyset$
    \FOR{each received update $(\Delta\theta^t_i, \sigma_i)$}
        \IF{$\texttt{verify}(\mathtt{pk}_i, \Delta\theta^t_i, \sigma_i)$}
            \STATE $S \leftarrow S \cup \{\Delta\theta^t_i\}$
        \ENDIF
    \ENDFOR
    \STATE $\theta^{t+1}_\mathtt{G} \leftarrow \theta^t_\mathtt{G} + \frac{1}{|S|} \sum_{\Delta\theta^t_i \in S} \Delta\theta^t_i$
\ENDFOR

\STATE \RETURN $\theta^T_\mathtt{G}$
\end{algorithmic}
\end{algorithm}

\section{Experiments}

\subsection{Implementation Details}

\begin{table}[htbp]
  \centering
  \caption{Hyper-parameters of the federated learning training for evaluation used in our work. We report the number of federated learning clients, total number of epochs, batch size, and learning rate~(LR).}
   \resizebox{0.9\linewidth}{!}{
    \begin{tabular}{ll|cccc}
    \toprule
    \midrule
    Model & Task & \# Clients & \# Epochs & Batch Size & LR \\
    \midrule
    MLP & MNIST & $10$ & $10$ & $32$ & $1e-5$ \\
    Pythia-$70$M & RTE & $10$ & $50$ & $8$ & $3e-5$ \\
    ResNet-$18$ & CIFAR-$100$ & $10$ & $20$ & $64$ & $3e-5$ \\
    \midrule
    \bottomrule
    \end{tabular}}
  \label{tab:hyper-parameters}
\end{table}

\paragraph{Models and Datasets} 
Our evaluation encompasses both vision and natural language processing tasks, representing two of the most significant domains in deep learning. We employ three diverse settings to assess the impact of PQC algorithms on federated learning across different scales and modalities:

\begin{itemize}
    \item \textbf{MLP on MNIST}: For a preliminary evaluation on a classic computer vision task, we use a Multi-Layer Perceptron (MLP) model on the MNIST dataset. MNIST consists of $70000$ $28\times28$ grayscale images of handwritten digits, providing a straightforward yet informative benchmark for image classification.
    \item \textbf{Pythia-$70$M on RTE}: To examine performance on language tasks, we employ the Pythia-70M model, a 70 million parameter language model from the Pythia series, on the Recognizing Textual Entailment|(RTE) dataset. RTE is a natural language inference task from the GLUE benchmark, challenging the model to determine the relationship between pairs of sentences.
    \item \textbf{ResNet-$18$ on CIFAR-$100$}: For a more complex vision task, we utilize ResNet-$18$, an $18$-layer residual neural network, on the CIFAR-100 dataset. CIFAR-$100$ comprises $60000$ $32\times32$ color images across $100$ classes, presenting a more challenging image classification scenario than MNIST.
\end{itemize}

\paragraph{Federated Learning Details}

Our experiments are built upon the \texttt{FedAvg} algorithm, a fundamental approach in federated learning that aggregates local model updates from distributed clients. While our implementation focuses on \texttt{FedAvg}, our proposed PQC-enhanced protocol is adaptable to most federated learning algorithms where communication occurs between clients and the central server. To ensure consistency and facilitate fair comparison across different PQC algorithms, we maintain uniform hyper-parameters for each client within a given model and task setting. The data distribution across clients follows an independent and identically distributed~(\textit{i.i.d}) splitting strategy, which allows us to focus on the impact of the PQC algorithms without the added complexity of non-\textit{i.i.d} data challenges. We provide a comprehensive list of the hyper-parameters used in our experiments in Table~\ref{tab:hyper-parameters}, including learning rates, batch sizes, and the number of local epochs for each setting.  We use NVIDIA H$100$ GPU and PyTorch to conduct our experiments. All models are trained with the constant learning rates and AdamW optimizer.

\begin{figure*}[htbp]
    \centering
    \includegraphics[width=1.0\linewidth]{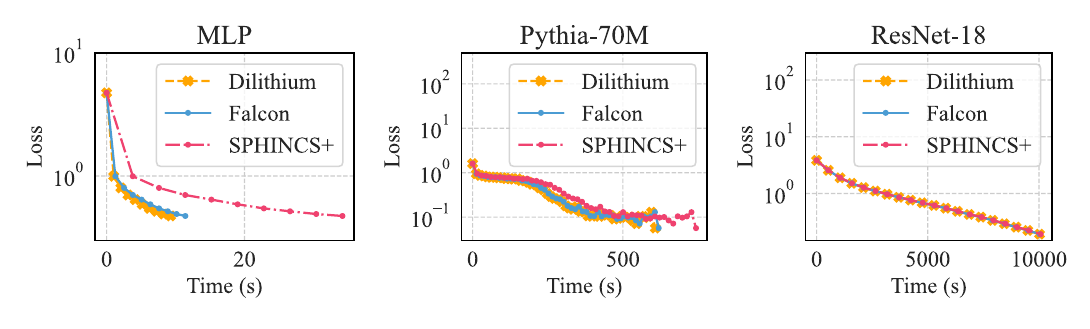}
    \caption{Federated learning training curves of the MLP on MNIST~(left), Pythia-$70$M on RTE~(middle), and the ResNet-$18$ on CIFAR-$100$. We evaluate the training time with the $3$ different PQC algorithms.} 
    \label{fig:main-results}
\end{figure*}

\subsection{Main Results}

Figure 3 presents the training curves for three different model-dataset combinations: MLP on MNIST, Pythia-$70$M on RTE, and ResNet-$18$ on CIFAR-$100$. Each plot shows the loss over time for the three PQC algorithms: Dilithium, Falcon, and SPHINCS+. These graphs allow us to compare both the convergence speed and final performance across different PQC implementations in various federated learning scenarios.

\paragraph{Training Speed Comparison} 
Our results clearly demonstrate a consistent ranking in terms of training speed across all three scenarios: SPHINCS+ is the slowest, followed by Falcon, with Dilithium being the fastest. This observation aligns with the characteristics summarized in Table~\ref{tab:algo-summary}. The impact of these speed differences for digital signature is less pronounced in the more complex models and datasets, particularly evident in the Pythia-$70$M and ResNet-$18$ experiments, where the local model training time scales extend to hundreds and thousands of seconds, respectively, being much higher than that of digital signature.

\paragraph{Consistency in Final Performance}
These results indicate that the choice of the PQC algorithm does not necessarily affect the final training performance, as measured by the loss values. Across all three scenarios, we observe that the loss curves converge to exact values regardless of the PQC algorithm used. This outcome is natural, as the PQC algorithms are designed to protect the integrity of model updates during communication without altering their content.

\paragraph{Trade-offs Between Security and Efficiency}
These results highlight the inherent trade-offs between security guarantees and computational efficiency in post-quantum federated learning. While SPHINCS+ offers strong security guarantees due to its hash-based nature, it consistently shows slower training times. Conversely, Dilithium balances strong security and computational efficiency, making it an attractive option for many federated learning scenarios. Falcon sits between these extremes, offering a middle ground in terms of both security and performance.

\subsection{Extended Study on Quantum Neural Network}

\begin{figure}[htbp]
    \centering
    \includegraphics[width=0.8\linewidth]{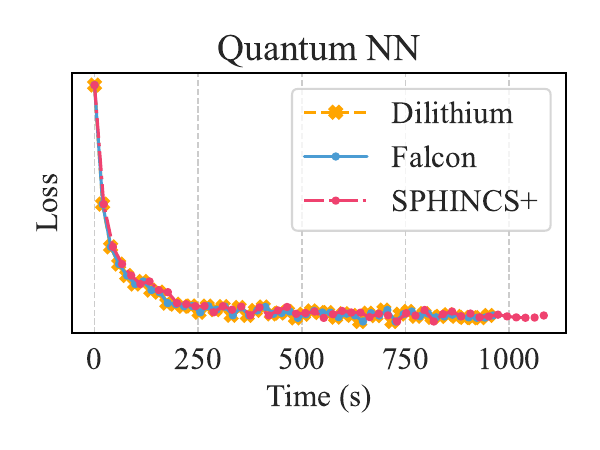}
    \caption{Federated learning training curves of the QNN model on MNIST. We evaluate the training time with the $3$ PQC algorithms.} 
    \label{fig:qnn-results}
\end{figure}

As we progress towards the post-quantum era, it is increasingly likely that we will train and deploy quantum neural networks (QNNs) on pure quantum computers. This shift necessitates an evaluation of PQC methods in the context of quantum neural network federated learning. To address this forward-looking scenario, we extended our study to include a simulated quantum neural network environment.

\paragraph{Experimental Setting} 
We implemented a federated learning system using a simulated quantum neural network. The simulation was conducted on classical GPUs, which, while not as efficient as a true quantum computer, allows us to approximate the behavior and challenges of quantum neural network training in a federated setting. We implement the training procedure with TorchQuantum\footnote{https://github.com/mit-han-lab/torchquantum}~\cite{hanruiwang2022quantumnas} on NVIDIA H$100$ GPU.

\paragraph{Results and Analysis}
Figure~\ref{fig:qnn-results} presents the training curves for the quantum neural network using the three PQC algorithms: Dilithium, Falcon, and SPHINCS+. The graph shows the loss over time for each algorithm during the federated learning process. Key observations are:

\begin{itemize}
    \item \textbf{Minimal Speed Differences}: Unlike our observations with classical MLP and Pythia-$70$M neural networks, the three PQC algorithms do not yield substantial differences in training speed for the quantum neural network. This is primarily attributed to the high computational cost of simulating quantum circuits on classical hardware, which dominates the overall training time.
    \item \textbf{Simulation Overhead}: The extended training time, which is up to $1000$ seconds, for this relatively simple quantum neural network underscores the current limitations of simulating quantum circuits on classical hardware. This overhead masks the performance differences between the PQC algorithms that were more apparent in classical neural network training.
\end{itemize}

\section{Discussion and Future Work}

\paragraph{Man-in-the-Middle Attack~(MITM)} While our proposed approach enhances the quantum security of federated learning using PQC digital signature algorithms, it is important to acknowledge a potential limitation in our current setting. Specifically, the system remains vulnerable to a sophisticated Man-in-the-Middle~(MITM) attack during the initial key exchange phase. If an adversary successfully intercepts and replaces the public keys during distribution, they could effectively compromise the entire system. In this scenario, the attacker could replace legitimate public keys with their own, allowing them to intercept, decrypt, and potentially modify all subsequent communications. This would enable the attacker to inject malicious model updates, effectively poisoning the global model without detection by the central server or other clients. This vulnerability underscores the critical importance of secure key distribution and verification mechanisms in federated learning systems, even when employing quantum-resistant cryptographic algorithms. Future research should focus on developing robust protocols for secure key exchange and distribution in distributed learning environments, possibly leveraging additional cryptographic primitives or trusted hardware solutions to mitigate this risk.

\paragraph{Evaluation of the Communication Time}
A significant limitation of our current study is the absence of a comprehensive evaluation of the communication costs associated with the three PQC signature algorithms in federated learning. Our experiments were conducted on a single GPU, which effectively eliminates real-world network communication overhead. This simplification, while allowing us to focus on computational performance, overlooks a crucial aspect of federated learning systems: the impact of signature size on communication efficiency. The three PQC algorithms - Dilithium, Falcon, and SPHINCS+ - have notably different signature sizes, which could significantly affect the overall system performance in a distributed setting. For instance, SPHINCS+ is known for its larger signature size compared to Dilithium and Falcon, which could potentially lead to increased communication overhead in a real-world federated learning deployment. Future work should address this gap by implementing these PQC algorithms in a truly distributed environment, measuring not only the computational time but also the time required for secure communication of model updates. This evaluation would provide a more holistic view of the trade-offs between security, computational efficiency, and communication overhead in post-quantum federated learning systems. Such insights would be invaluable for practitioners in choosing the most suitable PQC algorithm for their specific federated learning applications, considering both the computational resources and network conditions.

\section{Related Works}

\paragraph{Security Protection for Federated Learning} 
The integrity of federated learning systems faces significant challenges, particularly in the form of model-stealing attacks. Participants in the FL process may surreptitiously embed malicious functionalities into the shared global model. For instance, an image classification model could be manipulated to misclassify specific images based on attacker-defined criteria, or a text prediction model might be coerced into generating predetermined completions for certain prompts. To counter these threats, researchers have developed various protective measures. One notable approach involves implementing homomorphic encryption techniques~\cite{papernot2017semisupervisedknowledgetransferdeep}. This method safeguards user data by facilitating parameter exchanges within an encrypted environment. However, it's worth noting that this security enhancement comes at a cost: the necessity for encoding parameters prior to transmission and the subsequent exchange of public-private key pairs for decryption can significantly increase communication overhead.

\paragraph{Post-Quantum Cryptography for KEM}
Post-quantum key encapsulation mechanisms~(KEM) are crucial components in cryptographic systems designed to withstand attacks from quantum computers. Several promising KEM candidates have emerged through rigorous evaluation processes. Lattice-based schemes, such as CRYSTALS-Kyber, which leverages the Module Learning with Errors problem, offer a balanced security profile, performance, and key size~\cite{Banerjee2019SapphireAC}. Code-based KEMs provide alternative approaches, with BIKE utilizing quasi-cyclic moderate-density parity-check~\cite{nosouhi2022weakkeyanalysisbikepostquantum} codes and Classic McEliece employing Goppa codes~\cite{9773945}. These post-quantum KEMs are undergoing extensive research and development, with implementation efforts focusing on various platforms, including FPGA, ASIC, and RISC-V architectures~\cite{9605604,9748063}. Key areas of improvement include computational efficiency, power consumption reduction, and enhanced resistance to side-channel attacks~\cite{9974404}. As quantum computing technologies advance, these post-quantum KEMs are expected to secure future communication systems against emerging quantum threats, ensuring long-term data protection in an evolving cryptographic landscape~\cite{10089619}.

\section*{Acknowledgement}
We thank Kaushik Seshadreesan for his helpful discussions. JL is supported by his startup fund from the University of Pittsburgh.

% \clearpage

\bibliographystyle{IEEEtranS}
\bibliography{references}

\end{document}